\documentclass{myelsart}

\usepackage{epsfig}

\usepackage{amssymb}

\begin{document}

\begin{frontmatter}



\title{Nonmetallic thermal transport in low-dimensional proximity structures with 
partially preserved time-reversal symmetry in a magnetic field}


\author{G. Tkachov}

\address{Institute for Theoretical Physics, Regensburg University, 93040 Regensburg, Germany\\
Department of Physics, Lancaster University, Lancaster, LA1 4YB, UK\\
Institute for Radiophysics and Electronics NASU, Kharkov, 61085, Ukraine}


\begin{abstract}
Gapped excitation spectra of Andreev states 
are studied in one- and two-dimensional
(1D and 2D) normal systems in superconducting contacts subject to a parallel magnetic field. 
In the ballistic regime, a specific interplay between magnetic field spin splitting and 
the effect of a screening supercurrent is found to 
preserve time-reversal symmetry for certain groups of Andreev states remaining 
gapped despite the presense of the magnetic field. 
In 1D wires such states can lead to a fractional thermal magnetoconductance 
equal to half of the thermal conductance quantum. 
In 2D systems the thermal magnetoconductance is also predicted to 
remain suppressed well below the normal-state value in a wide range of magnetic fields.
\end{abstract}


\end{frontmatter}

\section{Introduction}
\label{one}

Recently, there has been extensive experimental and theoretical work aimed 
at understanding various microscopic manifestations of 
the mesoscopic-scale superconducting proximity effect in normal metal (N)-
superconductor (S) structures~\cite{Cast,Kroemer,vanWees,Green,Nitta,Petrashov,Pannetier,Anthore,Jo,Nazarov,Been,Lambert,Zaikin,Thornton}. On the normal side of such systems the superconducting correlations are maintained in the course of 
Andreev scattering~\cite{Andreev} during which a particle incident from the N region 
with energy below the superconducting gap energy $\Delta$
coherently evolves into a Fermi sea hole with the opposite 
spin that retraces the particle trajectory (time-reversed path) 
back to the normal bulk.
Due to elastic disorder in the N conductor or its restricted geometry, 
a large number of consecutive phase-preserving Andreev reflections
can occur at the NS interface making it effectively 
transparent to pair current. This gives rise to a veriety of phase-coherent 
phenomena such as, 
e.g. the zero-bias enhancement of the electron conductance~\cite{Cast,Kroemer,vanWees,Green,Nitta,Been,Imry} 
above the value predicted by the theory of a 
single-event scattering at an NS boundary~\cite{BTK},
the finite bias and magnetic field anomalies of the
phase-coherent conductance~\cite{Been,Poirier,Les,Yip},
the formation of a superconducting minigap in the quasiparticle density of states
in normal systems~\cite{Volkov,Golubov,Belzig,Altland},
Andreev edge states~\cite{Taka,Uhlisch,Zuelicke} 
and billiards~\cite{Kosztin,Ihra,Eroms,Jacquod,Cserti}.

While phase-coherent charge transport has been receiving considerable attention, 
heat conduction properties of mesoscopic 
proximity structures have been explored to a much lesser extent. 
A few theoretical papers have dealt with the thermal conductance 
of Josephson junctions~\cite{Kulik,Guttman,Zhao}, Andreev barriers and interferometers~\cite{Claughton,Bez,Kopnin}, quantum wires 
with proximity-induced superconductivity~\cite{Thermo}. 
Only recently, there has been a breakthrough in experiments on 
thermoelectric properties of small metallic NS hybrids~\cite{Chandra,Parsons}. 

The purpose of this paper is to report a theoretical study of anomalous magnetic field behaviour 
of heat transport in low-dimensional proximity structures. 
The choice of low-dimensional systems
is motivated by a progress in fabrication of superconducting contacts to 
high-mobility semiconductor quantum wells (see e.g. Refs.~\cite{Nano,Schaepers}). 
A number of experimental observations have suggested that due to the proximity effect 
such systems acquire properties of "clean" superconductors where the electron mean free
path $l$ is large compared to the induced superconducting coherence length $\xi_N$~\cite{Kroemer,Jo,Thornton,Nano,Schaepers}. 
More specifically, in planar semiconductor-superconductor junctions 
the proximity effect can be described in terms of Andreev 
bound states formed between the NS boundary and the back wall of the quantum well~\cite{Kroemer,Volkov}. 
Each Andreev state is a mixed particle-hole excitation whose spectrum, according to 
the theory of Ref.~\cite{Volkov}, should have a superconducting 
minigap $E_g$ smaller than $\Delta$ due to a residual interfacial barrier.
The existence of the minigap can for instance provide a natural explanation for the
unusually strong enhancement of the electric conductance reported in Refs.~\cite{Kroemer,Eroms}. 

The ballistic character of electron motion in semiconductor quantum wells 
needs to be taken into account when studying a magnetic field influence
on the proximity effect. As proposed in Ref.~\cite{Thermo}, 
a magnetic field ${\bf B}$ parallel to the plane of the quantum well [Fig.~\ref{Geo}(a)]
affects the Andreev states via a screening supercurrent induced in the superconductor.
In this case a finite Cooper pair momentum $2{\bf P}_S(B)$ at the NS boundary violates
the time-reversal conjugation of particles and holes which is accompanied by
a Galilean energy shift ${\bf p}{\bf P}_S(B)/m_N$ of the Andreev states 
($m_N$ and ${\bf p}$ are the electron mass and momentum in the plane of the quantum well). 
Since $P_S(B)\propto B$, the energy shift leads to the gapless excitation spectrum 
and ultimately to the metallic behaviour of the thermal conductance.

The above mechanism of the magnetic field influence on the Andreev states
neglects the energy of the Zeeman splitting $\alpha g\mu_B B$
(where $\alpha =\pm 1/2$, $g$ and $\mu_B$ are respectively the electron spin, g-factor and Bohr magneton). 
On the other hand, in InAs-based heterostructures which are commonly used
for contacts with superconductors
the g-factor can be as large as 10-13~\cite{Nano}, and therefore 
the magnetic spin splitting may play an important role. 
At first glance, the combined influence of the 
screening supercurrent and the spin splitting on the superconducting proximity
should anyway be a destructive one because both of them break time-reversal symmetry.
However, as shown below the interplay between the Zeeman energy $\alpha g\mu_B B$
and the Galilean shift ${\bf p}{\bf P}_S(B)/m_N$, which are both linear in $B$, can 
lead to the existence of Andreev states with the minigap 
independent of the magnetic field.
Unlike the case of spinless electrons~\cite{Thermo},
the thermal conductance remains anomalously small 
compared to that of a normal state even at relatively strong fields, 
up to the critical fields of the superconductor.

The predicted behaviour of the thermal magnetotransport
is also rather different from the observed 
in diffusive superconductors~\cite{Tinkham,AmbGrif,Maki} 
where the constructive interplay of the supercurrent and Zeeman effects 
is obstructed by strong momentum scattering
due to which the linear Galilean term averages out~\cite{Fulde}. 
Instead, the orbital magnetic field influence 
is described by an isotropic in momentum space depairing energy 
which is of higher order in $B$ 
(see e.g. Refs.~\cite{Maki,Deutscher,Anthore}).
Thus, the mechanism of the magnetic field influence on the Andreev states and their thermal conductance 
discussed in this paper is unique to ballistic proximity structures.

\section{Proximity effect on the excitation spectrum of ballistic electrons}
\label{two}

\begin{figure}[!t]
\begin{center}
\epsfxsize=0.4\hsize
\epsffile{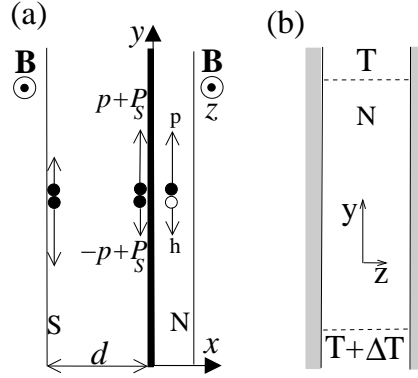}
\end{center}
\caption{
(a) Cross-sectional view of a superconductor (S)-normal system (N) junction.
The N system can be either a Q1D channel along the $y$-axis or a two-dimensional 
electron system located in the $y,z$ plane. 
The particle (p) and hole (h) momenta in the N system are both 
shifted by $P_S=(e/c)Bd/2$ 
in order to match the Cooper pair momentum $2P_S$ at the surface of the S film. 
(b) Model for studying ballistic thermal transport through 
a proximity-affected normal channel (N) connecting two reservoirs 
with temteratures $T$ and $T+\Delta T$ ($\Delta T\ll T$). 
}
\label{Geo}
\end{figure}

We first discuss the magnetic field influence on 
the proximity effect in a quasi-one-dimensional electron system (Q1DES) 
coupled in parallel to a superconducting film via a barrier of low transparency
$\tau$ as shown in Fig.~\ref{Geo}(a). The generalization for two-dimensional normal 
systems will be given in the next section.
The film thickness $d$ is assumed much smaller 
than both the superconducting coherence length and the London 
penetration depth in a parallel magnetic field ${\bf B}=[0; 0; B]$. 

As has been mentioned in the Introduction, in the geometry of Fig.~\ref{Geo}(a) 
the subgap states in the quantum well are formed in the course of 
multiple Andreev reflections which mix particles and holes producing a 
minigap in the excitation spectrum~\cite{Volkov}. For a weakly coupled
quantum well, the minigap energy $E_g\approx (v_N/v_S)\tau E_0$ depends on the energy
of the lowest occupied subband $E_0$, the interfacial transparency 
$\tau\ll 1$ and the ratio of the Fermi velocities in the normal ($v_N$) and 
superconducting ($v_S$) systems. In a narrow Q1D channel the motion of the mixed 
particle-hole excitations along the channel ($y$-direction in Fig.~\ref{Geo}(a)) 
can be desribed by two coupled equations for the annihilation and creation operators,
$\psi_{\alpha p}(t)$ and $\psi^\dag_{-\alpha -p}(t)$:

\begin{eqnarray}
&
\left[i\hbar\partial_t-\frac{(p+P_S)^{2}}{2m_N}+\alpha g\mu_B B+
E_N\right]\psi_{\alpha p}(t)=
E_gi\sigma_2^{\alpha\alpha^\prime}\psi^\dag_{\alpha^\prime -p}(t),&
\label{Eq1}\\
&
\left[-i\hbar\partial_t-\frac{(-p+P_S)^{2}}{2m_N}-\alpha g\mu_B B+
E_N\right]\psi^\dag_{-\alpha -p}(t)=
E_gi\sigma_2^{-\alpha\alpha^\prime}\psi_{\alpha^\prime p}(t),\quad&
\label{Eq2}\\
&
P_S=(e/c)Bd/2,&
\label{P}
\end{eqnarray}
%
where $E_gi\sigma_2^{\alpha\alpha^\prime}$ plays the role of the effective singlet pairing energy
($\sigma_2$ is the Pauli matrix),
$E_N$ stands for the Fermi energy in the Q1ES, 
and $p\equiv p_y$ is the quasiparticle momentum. 
The magnetic field influence on the quasiparticle spin is taken into account 
by the Zeeman term $\alpha g\mu_B B$, whereas 
the orbital effect is described by the shift $P_S$ of both particle and hole momenta
due to the screening supercurrent induced 
at the surface of the superconductor [see also Fig.~\ref{Geo}(a)]. 
As the thickness of the normal channel is considered negligible
compared to the superconductor thickness $d$, 
the shift of the electron and hole momenta in the Q1DES 
can be taken equal to the surface Cooper pair
momentum given per electron by Eq.~(\ref{P})  
($e>0$ is the absolute value of the electron charge). 
$P_S$ is proportional to the half-thickness of the superconductor
reflecting the fact that the field fully penetrates the film 
and generates an antisymmetric (linear) distribution of the supercurrent density 
with respect to its middle plane. 

One should note that the description of the proximity-induced correlations by coupled (superconductor-like) equations of motion (\ref{Eq1}) and (\ref{Eq2}) 
has to be reconciled with the fact that in the normal system there is no intrinsic superconducting
pairing. As shown in Appendix~\ref{Eqs}, the superconducting coupling in Eqs.~(\ref{Eq1}) and (\ref{Eq2})
is induced through the boundary conditions at the NS interface, 
for which the thickness of the quantum well 
must be of the order of the Fermi wavelength. 

The solution of equations (\ref{Eq1}) and (\ref{Eq2}) 
is given by the Bogolubov transformation of the form
\begin{eqnarray}
&&
\psi_{\alpha p}(t) 
= u_{p} b_{\alpha p}\exp(-it\epsilon_{\alpha p}^{+}/\hbar ) 
+i\sigma_2^{\alpha,-\alpha}
v_{p} b_{-\alpha -p}^\dag
\exp(-it\epsilon_{\alpha p}^{-}/\hbar ),
\label{sol}\\
&&
u_{p}^2 =\frac{1}{2}
\left[1+\frac{v_N(|p|-p_N)}{[v_N^2(|p|-p_N)^2+E_g^{2}]^{1/2}}
\right], 
v_{p}^2 =1-u_{p}^2, 
\nonumber
\end{eqnarray}
where $b_{\alpha p}$ and $b_{-\alpha -p}^\dag$ are Bogolubov's
quasiparticle operators, $p_N$ is the Fermi momentum and the excitation spectrum 
$\epsilon_{\alpha p}^{\pm}$ can be represented as 
\begin{eqnarray}
\epsilon_{\alpha p}^{\pm}&=&v_NP_S\,{\rm sgn}p -\alpha
g\mu_B B\pm[v_N^2(|p|-p_N)^2+E_g^{2}]^{1/2}=
\nonumber\\
&=&(k_Nd\,{\rm sgn}p -\alpha g)\mu_B B\pm[v_N^2(|p|-p_N)^2+E_g^{2}]^{1/2},
\label{Spectr}
\end{eqnarray}
with $k_N$ being the Fermi wave-number.
We neglect the term quadratic in $P_S$ assuming $P_S^2/2m_N\ll E_N$.

\begin{figure}[t]
\begin{center}
\epsfxsize=0.3\hsize
\epsffile{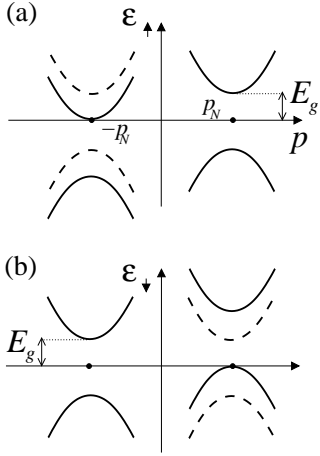}
\end{center}
\caption{\label{E}
Schematic view of the excitation spectrum (\ref{Spectr})
at finite magnetic fields ($B=B_g$): 
spin-up quasiparticles near $p=p_N$ (a) and spin-down 
ones near $p=-p_N$ (b) remain unaffected by the magnetic field 
due to the cancelation of the Zeeman and supercurrent effects 
at $g=2k_Nd$. Dashed curves correspond to $B=0$. 
}
\end{figure}

According to Eq.~(\ref{Spectr}) the magnetic field shifts the quasiparticle dispersion curves 
at the points $p=\pm p_N$ with respect to the Fermi level. At a certain field $B_g$ when $E_g=(g/2+k_Nd)\mu_B B_g$ there appear first gapless excitations 
near the Fermi points: spin-up ones at $p=-p_N$ and spin-down ones at $p=p_N$ as sketched 
in Fig.~\ref{E}. 
This field can be expressed in terms of the flux quantum $\Phi_0$, proximity-induced
coherence length $\xi_N=\hbar v_N/2E_g$ and superconductor thickness $d$ as follows:
\begin{eqnarray}
B_g=\frac{\Phi_0}{\pi\xi_N d(1+\tilde g)}, \qquad\qquad \tilde g=\frac{g}{2k_Nd}.
\label{Bg}
\end{eqnarray}
Alternatively, it can be related to the critical fields of the superconducting
film as:
\begin{eqnarray}
B_g=B_{orb}\frac{\lambda}{\xi_N(1+\tilde g)}=B_{spin}\frac{E_g}{\Delta k_Nd(1+\tilde g)},
\label{Bg1}
\end{eqnarray}
where $B_{orb}\approx\Phi_0/\lambda d$ is the critical field due to the orbital effect
of the magnetic field ($\lambda$ is the London penetration depth) 
and $B_{spin}\approx \Delta/\mu_B$ is the critical field of the paramagnetic 
limit~\cite{Spin}. According to Eqs.~(\ref{Bg1}) for junctions with well separated
gap energies $E_g\ll\Delta$ and $k_Nd\sim g>1$ the field $B_g$ is much smaller 
than both critical fields.

It follows from Eq.~(\ref{Spectr})
that for the special case where $g=2k_Nd$   
the Zeeman and supercurrent terms cancel each other at one 
of the Fermi points
and the minigap remains at the Fermi level [see Fig.~\ref{E}]. 
For these excitations time-reversal symmetry is 
exactly preserved despite the presence of the magnetic field. 
If $g\not= 2k_Nd$, the excitations near both Fermi points eventually become gapless 
as the magnetic field increases. 
However in a finite range of fields given by 
\begin{equation}
1<\frac{B}{B_g}<\frac{1+\tilde g}{|1-\tilde g|},
\label{range}
\end{equation}
the minigap still exists at one of the Fermi points, which accounts for the anomalous behaviour
of the thermal conductance of low-dimensional proximity systems discussed in the next
section. 

\section{Thermal conductance of low-dimensional proximity structures}
\label{three}

Since the pioneering works~\cite{Tinkham,Tcond} 
studies of heat transport 
have been playing an important role in understanding 
spectral properties of low-energy excitations in superconductors (see e.g.
Refs.~\cite{AmbGrif,Maki}).
In what follows we discuss how the unusual magnetic field behaviour of 
the quasiparticle energies in low-dimensional proximity superconductors reflect on 
their thermal conductance characteristics.

\subsection{Q1D channel}

One more advantage of describing the proximity effect in the Q1DES by 
the superconductor-like operator equations (\ref{Eq1}) and (\ref{Eq2}) 
is that we can use the well known procedure of Ref.~\cite{KadMar} to calculate
the heat current $j_Q$. 
For a long ballistic Q1D electron channel 
between two reservoirs with temperatures $T$ and $T+\Delta T$ $(\Delta T\ll T)$ [see Fig.~\ref{Geo}(b)], 
the formal derivation of $j_Q$ was given in Ref.~\cite{Thermo}.
To discuss the combined influence of the Zeeman spin splitting and the supercurrent 
on thermal transport in the channel we can start with the following 
expression for $j_Q$~\cite{Thermo}:

\begin{eqnarray}
j_Q=h^{-1}\sum_{\alpha}
& &
\left[ 
\int\limits_{p\leq p_N}dp\,
\epsilon_{\alpha p}^{+}
\partial_{p}\epsilon_{\alpha p}^{+}n_<(\epsilon_{\alpha p}^{+})+
\int\limits_{p\geq p_N}dp\,
\epsilon_{\alpha p}^{+}
\partial_{p}\epsilon_{\alpha p}^{+}n_>(\epsilon_{\alpha p}^{+})
+
\right.
\nonumber\\
& &
\left. 
\int\limits_{p\leq -p_N}dp\,
\epsilon_{\alpha p}^{+}
\partial_{p}\epsilon_{\alpha p}^{+}n_<(\epsilon_{\alpha p}^{+})+
\int\limits_{p\geq -p_N}dp\,
\epsilon_{\alpha p}^{+}
\partial_{p}\epsilon_{\alpha p}^{+}n_>(\epsilon_{\alpha p}^{+})
\right].
\label{j}
\end{eqnarray}
It is related to the excitation energies 
$\epsilon_{\alpha p}^{+}$ and 
the group velocities $\partial_{p}\epsilon_{\alpha p}^{+}$ 
of the "+" branch of the quasiparticle spectrum (\ref{Spectr}).
We have taken into account the contribution of the "-" branch by exploiting 
the symmetry relation $\epsilon_{\alpha p}^{-}=-\epsilon_{-\alpha -p}^{+}$ 
between the two spectrum branches.

The four terms in Eq.~(\ref{j}) correspond to the two 
rightmoving ($\partial_{p}\epsilon_{\alpha p}^{+}>0$) and two 
leftmoving ($\partial_{p}\epsilon_{\alpha p}^{+}<0$) quasiparticle 
modes of the excitation spectrum shown in Fig.~\ref{E}.
The distribution functions of the rightmovers 
and the leftmovers 
are assumed to be set by the reservoirs, 
as 
$n_>(\epsilon_{\alpha p}^{+})=
n(\epsilon_{\alpha p}^{+},T+\Delta T)$ and 
$n_<(\epsilon_{\alpha p}^{+})=
n(\epsilon_{\alpha p}^{+},T)$, respectively, 
with $n(\epsilon_{\alpha p}^{+},T)$ being the Fermi function.
The current (\ref{j}) can now be expressed in terms of the 
differences $n_>(\epsilon_{\alpha p}^{+})-n_<(\epsilon_{\alpha p}^{+})$,
and finally the thermal conductance can be introduced 
as the proportionality coefficient between the heat current and the temperature drop, 
$j_Q=\kappa_1(B,T)\Delta T$, where for $\kappa_1(B,T)$ we have

\begin{eqnarray}
\frac{\kappa_1(B,T)}{\kappa_{1N}(T)}=
\frac{3}{\pi^2}\sum_{\alpha}
\left[
\int\limits_{\frac{E_g-(k_Nd+\alpha g)\mu_B B}{2k_BT}}^{\infty}
\frac{x^2dx}{\cosh^2 x}
+
\int\limits_{\frac{E_g+(k_Nd-\alpha g)\mu_B B}{2k_BT}}^{\infty}
\frac{x^2dx}{\cosh^2 x}  
\right].
\label{k}
\end{eqnarray}
Here $\kappa_{1N}(T)=\pi k_B^2T/3\hbar$ is the thermal conductance of a normal channel~\cite{Pendry} 
($k_B$ is the Boltzmann constant).
It is convenient to rewrite equation (\ref{k}) in terms of the dimensionless magnetic 
field $b=B/B_g$, temperature $t=k_BT/E_g$ and the ratio of the Zeeman 
and supercurrent energies $\tilde g$ [see equation (\ref{Bg})] as:

\begin{eqnarray}
\frac{\kappa_1(b,t,\tilde g)}{\kappa_{1N}}=
\frac{3}{\pi^2}
\left[
\int\limits_{\frac{1-b}{2t}}^{\infty}
\frac{x^2dx}{\cosh^2 x}\right.
+
\int\limits_{\frac{1-(\tilde g-1)b/(\tilde g+1)}{2t}}^{\infty}
\frac{x^2dx}{\cosh^2 x}
&+&
\nonumber\\
\int\limits_{\frac{1-(1-\tilde g)b/(1+\tilde g)}{2t}}^{\infty}
\frac{x^2dx}{\cosh^2 x}
&+&
\left.
\int\limits_{\frac{1+b}{2t}}^{\infty}
\frac{x^2dx}{\cosh^2 x}
\right].
\label{k1}
\end{eqnarray}
\begin{figure}[t]
\begin{center}
\epsfxsize=0.8\hsize
\epsffile{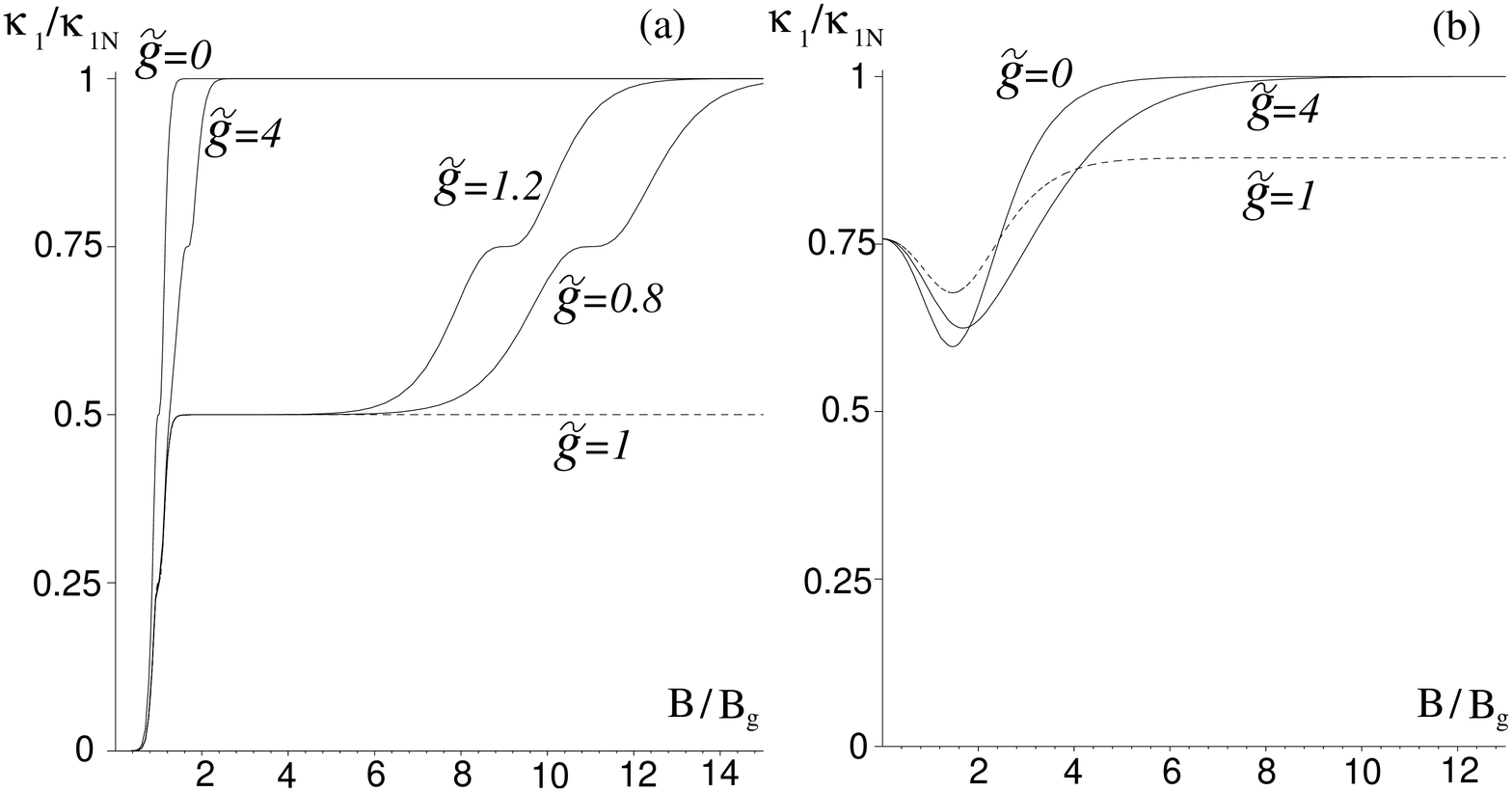}
\end{center}
\caption{\label{k_B}
Magnetic field dependence of the normalized thermal conductance for (a) $k_BT/E_g=0.05$ and 
(b) $k_BT/E_g=0.5$.}
\end{figure}

In the absence of the magnetic field ($b=0$) the conductance (\ref{k1})
is exponentially small at temperatures below the minigap ($t<1$). 
As shown in Fig.~\ref{k_B}(a) low-temperature thermal transport 
can be stimulated by applying a magnetic field above the threshold (\ref{Bg}) 
corresponding to the appearance of gapless excitations.
However the character of the magnetic field behaviour of the thermal conductance crucially 
depends on the ratio of the Zeeman and supercurrent energies $\tilde g$.
For $\tilde g=1$ the conductance saturates at half of the normal metallic value
$\kappa_{1N}/2$ in contrast with both cases of small and large $\tilde g$ 
where it fully recovers the metallic behaviour. This difference results
from the compensation of the Zeeman and supercurrent effects for half 
of the excitations which remain gapped for $\tilde g=1$ despite the presence 
of the magnetic field [see Fig.~\ref{E}]. Figure~\ref{k_B}(a) also shows 
that any small deviation from $\tilde g=1$ eventually drives the system to 
the metallic regime. In this case the anomalous "half-metallic" regime 
is present in a finite range of intermediate magnetic fields given by equation (\ref{range}).
If $\tilde g$ is very close to 1, the upper limit of this range 
$B=B_g (1+\tilde g)/|1-\tilde g|$ can approach the lowest of 
the critical fields of the superconducting film, $B_{orb}$ or $B_{spin}$. 
This means that despite small values of $E_g$
compared to the superconducting gap energy $\Delta$ the proximity-induced thermal transport 
anomalies in one-dimensional systems 
can persist untill the order parameter in the superconductor is destroyed by the magnetic 
field.

\begin{figure}[t]
\begin{center}
\epsfxsize=0.35\hsize
\epsffile{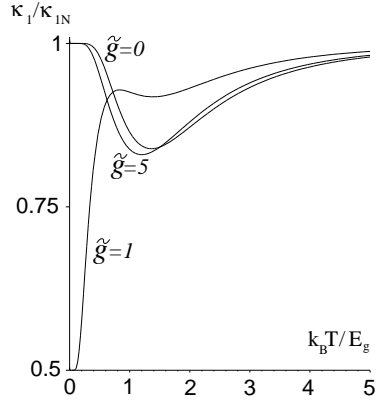}
\end{center}
\caption{\label{k_T}
Normalized thermal conductance vs temperature for $b=5$.}
\end{figure}

Figure~\ref{k_B}(b) demonstrates that the anomalous magnetic field behaviour 
of the thermal conductance can still be well distinguished at temperatures of order of 
$E_g/k_B$. This can be directly seen from comparing 
the temperature dependences for $\tilde g=1$ and $\tilde g=0$ (or $\tilde g>1$)
as shown in Fig.~\ref{k_T}.

\begin{figure}[b]
\begin{center}
\epsfxsize=0.8\hsize
\epsffile{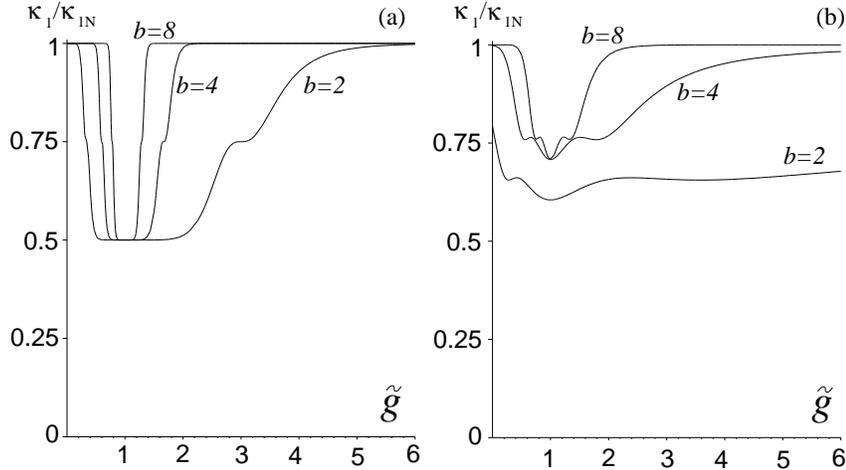}
\end{center}
\caption{\label{k_g}
Normalized thermal conductance vs ratio of the Zeeman and supercurrent energies 
${\tilde g}=g/2k_Nd$ for (a) $k_BT/E_g=0.05$ and 
(b) $k_BT/E_g=0.3$. 
Such a dependence can result from a gate voltage variation of the electron $g$-factor or
the carrier concentration (i.e. $k_N$).}
\end{figure}

The parameter $\tilde g$ in Eq.~(\ref{Bg}) characterizing the relative strength of 
the Zeeman and supercurrent effects involves the superconductor thickness $d$, 
electron $g$-factor and the Fermi wave-number $k_N$. 
If the Q1DES is formed in a semiconductor 
heterostructure, the latter two can be controlled by an external gate 
voltage. This in principle gives one more way of manipulating thermal transport
in low-dimensional proximity structures. To investigate such a possibility,
in Fig.~\ref{k_g} we present the dependence of the normalized conductance 
on the parameter $\tilde g$. 
The surviving proximity effect manifests itself as 
a minimum of the ratio $\kappa_1(\tilde g)/\kappa_{1N}$ 
around the critical point $\tilde g=1$. This minimum reaches $1/2$ 
at low temperatures [Fig.~\ref{k_g}(a)] and remains well pronounced 
as the temperature approaches $E_g/k_B$ [Fig.~\ref{k_g}(b)].

\subsection{2D electron system}

Here we extend our approach to planar structures combining a superconductor
and a two-dimensional electron system (2DES). 
The 2DES is chosen to be located in the $y,z$ plane [Fig.~\ref{Geo}] and 
has a form of a strip of width $W$ (in the $z$-direction)
with a large number of the electron channels: $2W p_N/h\gg 1$. 
The supercurrent flows along the strip (in the $y$-direction).
The straightforward generalization of the excitation spectrum (\ref{Spectr}) 
for a 2D isotropic Fermi surface is

\begin{eqnarray}
\epsilon_{\alpha p_0\theta}^\pm=(k_Nd\cos\theta -\alpha g)\mu_B B 
\pm[v_N^2(p_0-p_N)^2+E_g^2]^{\frac{1}{2}},
\label{Spectr2}
\end{eqnarray}
where $p_0$ is the absolute value of the quasiparticle momentum ($p_0\approx p_N$),
$\theta$ is the angle between the quasiparticle momentum 
${\bf p}=[p_0\cos\theta; p_0\sin\theta]$ and that of a Cooper pair 
${\bf P}_S=[P_S; 0]$ in the 2DES plane. 
Note that in the direction of the supercurrent flow ($\theta=0$) and in the 
opposite direction ($\theta=\pi$) the quasiparticle dispersion 
(\ref{Spectr2}) looks like that sketched in Fig.~\ref{E}
near the Fermi points $p_0=p_N$ and $p_0=-p_N$, respectively.

We assume that the temperature drop is created in the direction of 
the supercurrent flow, i.e. along the strip [Fig.~\ref{Geo}(b)]. 
In this case the contribution of one electron channel to the net heat current 
is given by the 1D ballistic formula (\ref{j}). In the polar coordinates
$p_0,\theta$ the net heat current can be written as

\begin{eqnarray}
&
J_Q=\frac{2p_NW}{h^2}\sum_{\alpha}
\int\limits_0^{\pi/2}d\theta
\left[ 
\int\limits_{p_0\leq p_N}dp_0\,
\epsilon_{\alpha p_0\theta}^{+}\,
n_<(\epsilon_{\alpha p_0\theta}^{+})\,
\partial_{p_y}\epsilon^{+}_{\alpha}|_{p_0\theta}
+
\right. &
\nonumber\\
&
\qquad\qquad\qquad\qquad
\left.
+
\int\limits_{p_0\geq p_N}dp_0\,
\epsilon_{\alpha p_0\theta}^{+}\,
n_>(\epsilon_{\alpha p_0\theta}^{+})\,
\partial_{p_y}\epsilon^{+}_{\alpha}|_{p_0\theta}
\right]+&
\nonumber\\
&
+
\frac{2p_NW}{h^2}\sum_{\alpha}
\int\limits_{\pi/2}^{\pi}d\theta
\left[ 
\int\limits_{p_0\leq p_N}dp_0\,
\epsilon_{\alpha p_0\theta}^{+}\,
n_>(\epsilon_{\alpha p_0\theta}^{+})\,
\partial_{p_y}\epsilon^{+}_{\alpha}|_{p_0\theta}
+
\right.
&
\nonumber\\
&
\qquad\qquad\qquad\qquad
\left.
+
\int\limits_{p_0\geq p_N}dp_0\,
\epsilon_{\alpha p_0\theta}^{+}\,
n_<(\epsilon_{\alpha p_0\theta}^{+})\,
\partial_{p_y}\epsilon^{+}_{\alpha}|_{p_0\theta}
\right],
&
\label{J}
\end{eqnarray}
where the integrals over the angles $0\leq\theta\leq\pi/2$ and $\pi/2\leq\theta\leq\pi$ take 
into account the contributions of the channels with positive and negative 
$p_y$, respectively;
$\partial_{p_y}\epsilon^{+}_{\alpha}|_{p_0\theta}$ is the quasiparticle velocity in the 
direction of the temperature drop which has to be expressed in terms of the 
polar variables $p_0,\theta$. Note that with the assumed accuracy $|P_S|\ll p_N$ the 
supercurrent term in Eq.~(\ref{Spectr2}) does not affect the position 
of the minimum ($p_0\approx p_N$) of the quasiparticle energy $\epsilon^{+}_{\alpha p_0\theta}$. 
Calculating with the same accuracy the quasiparticle velocity, one finds  
$\partial_{p_y}\epsilon^{+}_{\alpha}|_{p_0\theta}
\approx \partial_{p_0}\epsilon^{+}_{\alpha p_0\theta}\cos\theta$.

As in the 1D case, to model ballistic non-equilibrium transport 
between two reservoirs with temperatures $T+\Delta T$ and $T$,
in Eq.~(\ref{J}) we introduce different equilibrium distribution 
functions $n_>(\epsilon_{\alpha p}^{+})=
n(\epsilon_{\alpha p}^{+},T+\Delta T)$ 
and
$n_<(\epsilon_{\alpha p}^{+})=
n(\epsilon_{\alpha p}^{+},T)$ for the quasiparticles with positive ($\partial_{p_y}\epsilon^{+}_{\alpha}|_{p_0\theta}>0$)
and negative ($\partial_{p_y}\epsilon^{+}_{\alpha}|_{p_0\theta}<0$)
velocities, respectively. Now the integrals over the momentum $p_0$
can be transformed into the energy integration as follows:

\begin{eqnarray}
&
J_Q=\frac{2p_NW}{h^2}\sum_{\alpha}
\int\limits_0^{\pi/2}\cos\theta d\theta
\left[ 
\int\limits_{E_g+(k_Nd\cos\theta -\alpha g)\mu_B B}^\infty d\epsilon\,
\epsilon\,[n_>(\epsilon)-n_<(\epsilon)]
\right]-&
\nonumber\\
&
-
\frac{2p_NW}{h^2}\sum_{\alpha}
\int\limits_{\pi/2}^{\pi}\cos\theta d\theta
\left[ 
\int\limits_{E_g+(k_Nd\cos\theta -\alpha g)\mu_B B}^\infty d\epsilon\,
\epsilon\,[n_>(\epsilon)-n_<(\epsilon)]
\right],
&
\label{J1}
\end{eqnarray}
and the conductance $\kappa_2(B,T)$ of the 2DES is introduced as 
$J_Q=\kappa_2(B,T)\Delta T$, where

\begin{eqnarray}
\frac{\kappa_2(B,T)}{\kappa_{2N}(T)}&=&
\frac{3}{\pi^2}\sum_{\alpha}\int_0^{\pi/2}\cos\theta d\theta\times
\nonumber\\
&\times&
\left[
\int\limits_{\frac{E_g-(k_Nd\cos\theta+\alpha g)\mu_B B}{2k_BT}}^{\infty}
\frac{x^2dx}{\cosh^2 x}
+
\int\limits_{\frac{E_g+(k_Nd\cos\theta-\alpha g)\mu_B B}{2k_BT}}^{\infty}
\frac{x^2dx}{\cosh^2 x}  
\right].
\nonumber
\end{eqnarray}
After exchanging the integrals over $\theta$ and $x$, the angle integral
can be easily calculated:
\begin{eqnarray}
\frac{\kappa_2(b,t,{\tilde g})}{\kappa_{2N}}&=&\frac{3}{\pi ^{2}}
\sum_{\alpha}
\left[
\int\limits_{\frac{1-2\alpha {\tilde g}b/({\tilde g}+1)}{2t}}^{\infty}
\frac{2x^2dx}{\cosh^2 x}
+
\right.
\label{k2}\\
&+&
\int\limits_{\frac{1-(2\alpha {\tilde g}+1)b/({\tilde g}+1)}{2t}}
^{\frac{1-2\alpha {\tilde g}b/({\tilde g}+1)}{2t}}
\frac{x^2dx}{\cosh^2 x}
\left(1-\frac{\left[(2tx-1)({\tilde g}+1)+2\alpha {\tilde g}b\right]^2}{b^2}\right)^{1/2}+  
\nonumber\\
&-&
\left.
\int\limits_{\frac{1-2\alpha {\tilde g}b/({\tilde g}+1)}{2t}}
^{\frac{1-(2\alpha {\tilde g}-1)b/({\tilde g}+1)}{2t}}
\frac{x^2dx}{\cosh^2 x}
\left(1-\frac{\left[(2tx-1)({\tilde g}+1)+2\alpha {\tilde g}b\right]^2}{b^2}\right)^{1/2}
\right],
\nonumber
\end{eqnarray}
where we again use the dimensionless magnetic field $b$, temperature $t$ 
and the parameter ${\tilde g}$ characterizing the relative strength of the Zeeman and 
supercurrent effects; $\kappa_{2N}(T)=(2p_NW/h)\kappa_{1N}(T)$ 
is the thermal conductance of a normal 2D stripe. In principle, the 2D geometry
allows for an arbitrary relative orientation of 
the heat current and the supercurrent, and in this case the thermal conductance will
depend on the angle between them. Our geometry where this angle is zero has an 
advantage of being equally suitable for studying heat transport in both Q1D and 2D systems.

\begin{figure}[t]
\begin{center}
\epsfxsize=0.8\hsize
\epsffile{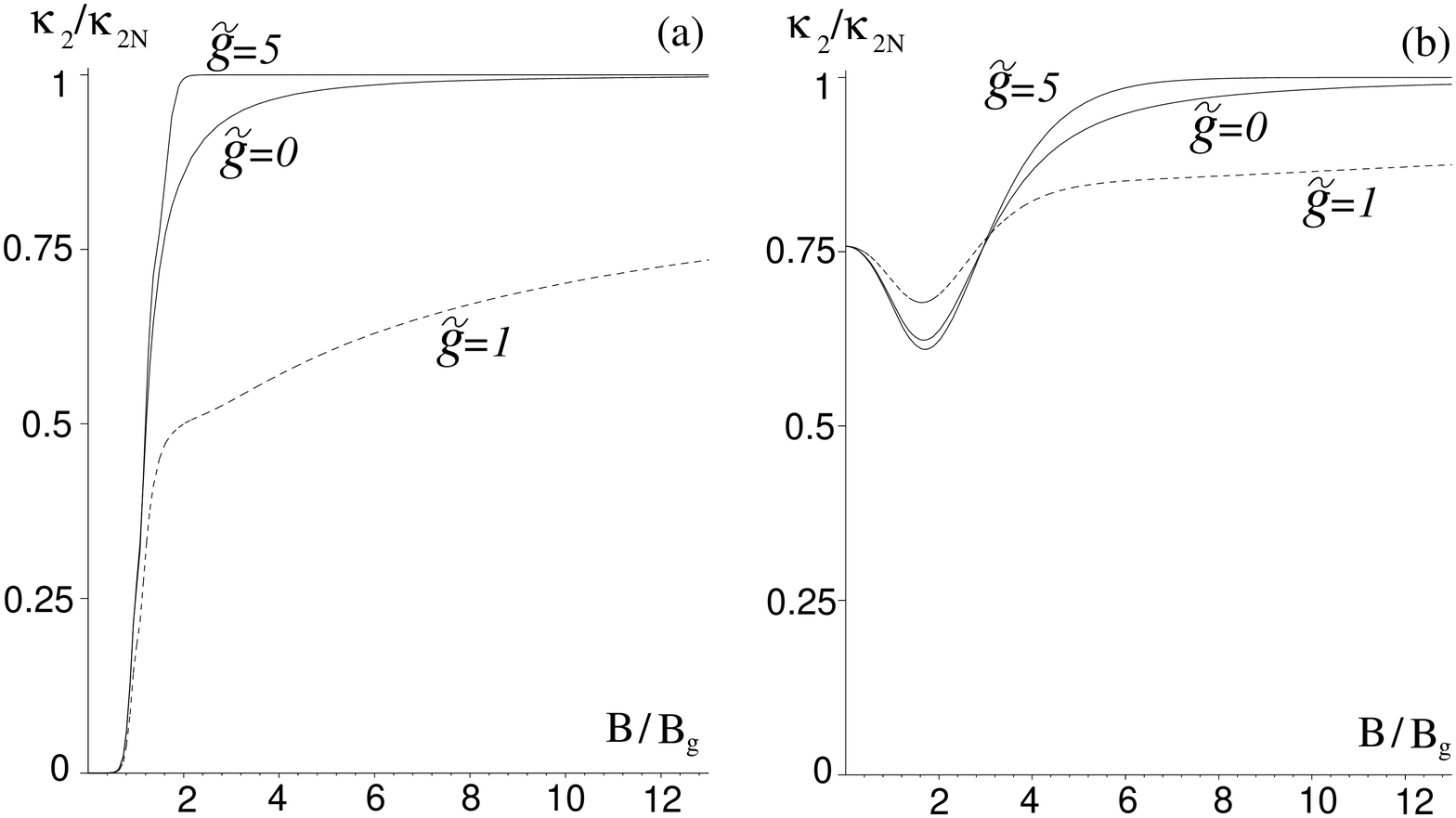}
\end{center}
\caption{\label{k2_B}
Normalized thermal conductance of a 2DES vs magnetic field for (a) $k_BT/E_g=0.05$ and 
(b) $k_BT/E_g=0.5$.}
\end{figure}

\begin{figure}[t]
\begin{center}
\epsfxsize=0.8\hsize
\epsffile{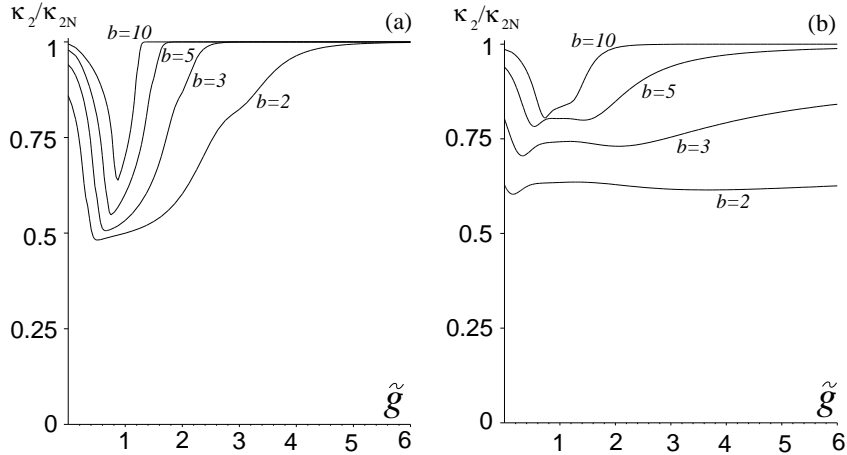}
\end{center}
\caption{\label{k2_g}
Normalized thermal conductance of a 2DES vs parameter ${\tilde g}=g/2k_Nd$ for (a) $k_BT/E_g=0.05$ and 
(b) $k_BT/E_g=0.4$.}
\end{figure}

Figure~\ref{k2_B}(a) shows the magnetic field dependence of the normalized conductance
(\ref{k2}) for different values of ${\tilde g}$. Like in the 1D case
for small and large values of ${\tilde g}$ the conductance quickly approaches 
its normal value when $b$ exceeds 1. However for ${\tilde g}=1$ 
this increase is much slower due to the compensation of the Zeeman and supercurrent effects
for spin-up quasiparticles moving along the supercurrent flow ($\theta =0$)
and for spin-down ones moving in the opposite direction ($\theta =\pi$).
There is no saturation in this case because the majority of the excitations 
(i.e. with the intermediate angles $\theta\not =0,\pi$) are still affected 
by the magnetic field: they all become gradually gapless as the field increases. 
The nonmonotonic dependence of the normalized conductance on the 
parameter ${\tilde g}$ shown in Fig.~\ref{k2_g} also suggests 
that for ${\tilde g}\approx 1$ the proximity effect in the 2DES survives in higher 
magnetic fields.

\section{Conclusion}
\label{four}

An intermediate regime, between superconducting and metallic, of heat transport 
has been predicted in ballistic low-dimensional proximity superconductors 
subject to a parallel magnetic field. 
It is due to the combined influence of the screening supercurrent
and the Zeeman spin splitting on the proximity-induced gap (minigap) 
in the quasiparticle states. 
The interplay of the two effects 
is found to preserve the minigap at the Fermi energy 
for certain groups of the quasiparticles despite the presence of the magnetic field. 
In this regime the low-temperature thermal 
conductance is about two times smaller than in the normal state and  
nearly independent of the magnetic field. For comparison, 
in conventional superconductors 
with a supercurrent~\cite{Maki,Fulde,Anthore} 
or the spin splitting~\cite{Maki,Spin} acting separately, 
no gapped excitations remain at the Fermi energy as the field increases, 
and the thermal conductivity fully recovers the metallic behaviour.

To observe the anomalous behaviour of the thermal conductance
the parameter $g/2k_Nd$ controlling the relative strength of 
the Zeeman and supercurrent effects must be close to 1.
This requirement can be met in nanostructures combining large g-factor semiconductor 
quantum wells (e.g. InAs-based) and thin superconducting films with typical 
thicknesses $d\sim k_N^{-1}\sim 10$ nm. It is important to emphasize 
that the magnetic field $B_g$ [Eq.~(\ref{Bg1})]
characterizing both supercurrent shift and Zeeman splitting of  
the Andreev states is small compared to the critical fields of
the superconducting film (arising from the orbital or paramagnetic effects).  
For this reason, we did not consider the magnetic field dependence of 
the order parameter in the superconductor and the effective pairing energy $E_g$ 
in the normal system. This question 
as well as the role of elastic scattering could be a subject of further studies.  

The author thanks V.~I.~Fal'ko for discussions. 
This work was supported by EPSRC (UK) and DFG (Germany).

\appendix

\section{Effective pairing energy description of 
the superconducting proximity effect in low-dimensional systems}
\label{Eqs}

Below we present the microscopic derivation of the equations of motion~(\ref{Eq1}) and (\ref{Eq2}) 
and discuss their applicability.
The approach used here is close in spirit to the tunneling self-energy description of the superconducting 
proximity effect proposed in Ref.~\cite{Mac}. 
However we will not employ the tunneling Hamiltonian formalism~\cite{Cohen} whose  
validity for high-order tunneling processes was a subject of debates~\cite{Prange,Caroli,Feucht}. 
Instead, to generate the tunneling self-energies we will use directly the equations of motion 
in the superconductor in the position representation. It turns out that in this approach there is no need 
to resort to the model of a homogeneous pairing potential in the superconductor. In this sense,
a more general treatment of the proximity effect can be given and the results 
can then be compared with those obtained for the homogeneous pairing~\cite{Volkov}.  

We assume a rectangular potential barrier with flat walls occupying the 
region $-a\leq x\leq 0$ which separates a superconductor ($x<-a$)
and a normal system ($x>0$) [see Fig.~\ref{Geo}(a) where the barrier
is shown as a thin black rectangle]. The Fermi energies 
$E_{S,N}$, Fermi momenta $p_{S,N}$ and the electron masses 
$m_{S,N}$ in the S and N systems are considered different. 
For simplicity, the derivation will be given for zero magnetic field. 

The superconductor is desribed by the mean-field equations of motion for the creation 
$\psi^\dag_{\alpha-{\bf p}}(x)$ and annihilation $\psi_{\alpha{\bf p}}(x)$ 
operators~\cite{Gorkov,deGennes}:

\begin{eqnarray}
&&
\left[i\hbar\partial_t-\varepsilon_S({\hat p}_x,{\bf p})
+E_S
\right]
\psi_{\alpha {\bf p}}(x,t)+
\Delta_{\alpha^\prime\alpha}(x)
\psi^\dag_{\alpha^\prime -{\bf p}}(x,t)=0,\label{EqS}\\
&&
\left[-i\hbar\partial_t-\varepsilon_S(-{\hat p}_x,-{\bf p})
+E_S
\right]
\psi^\dag_{\alpha -{\bf p}}(x,t)+
\Delta^*_{\alpha^\prime\alpha}(x)
\psi_{\alpha^\prime {\bf p}}(x,t)=0,
\nonumber
\end{eqnarray}
where $\Delta_{\alpha^\prime\alpha}(x)$ is the pairing potential, 
$\varepsilon_S({\hat p}_x,{\bf p})$ is the one-particle operator of the kinetic energy 
with ${\bf p}=[p_y, p_z]$ being the parallel (conserved) component of the electron 
momentum and ${\hat p}_x=-i\hbar\partial_x$, and we assume a summation over the spin projection $\alpha^\prime$. In the normal system (initially three-dimensional) the equations of motion are 
\begin{eqnarray}
&&
\left[i\hbar\partial_t-\varepsilon_N({\hat p}_x,{\bf p})-V(x)+E_N\right]
\psi_{\alpha {\bf p}}(x,t)=0,
\label{EqN}\\
&&
\left[-i\hbar\partial_t-\varepsilon_N(-{\hat p}_x,-{\bf p})-V(x)+E_N\right]
\psi^\dag_{\alpha -{\bf p}}(x,t)=0, 
\nonumber
\end{eqnarray}
where $\varepsilon_N({\hat p}_x,{\bf p})$ is the operator of the kinetic energy of a normal electron, 
$V(x)$ is a confining potential. 
Inside a high enough barrier 
one can neglect the energy and momentum dependence of the electron penetration length 
and write the equation of motion as:
$
[\partial^2_x-q^2]\chi_{\alpha{\bf p}}(x)=0,
$
where $q^{-1}=\hbar/(2m_BU)^{1/2}$ is the electron penetration length depending on 
the barrier height $U$ and the electron effective mass $m_B$. 
We introduce a special notation $\chi_{\alpha{\bf p}}(x)$ for the 
electron operator inside the barrier.
The continuity of the particle current imposes usual boundary conditions at the barrier walls: 
\begin{eqnarray}
  & 
  \chi_{\alpha{\bf p}}(0)=\psi_{\alpha{\bf p}}(0),\quad
  \chi_{\alpha{\bf p}}(-a)=\psi_{\alpha{\bf p}}(-a),&
  \label{B-bound}\\
  &
  \partial_x\chi_{\alpha{\bf p}}(0)=\frac{m_B}{m_N}
  \partial_x\psi_{\alpha{\bf p}}(0),
  \qquad
  \partial_x\chi_{\alpha{\bf p}}(-a)=\frac{m_B}{m_S}
  \partial_x\psi_{\alpha{\bf p}}(-a).&
  \label{B-derbound}
\end{eqnarray}

The solution inside the barrier satisfying boundary conditions (\ref{B-bound}) is 

\begin{eqnarray}
\chi_{\alpha{\bf p}}(x)
=
\frac{\sinh q(x+a)}{\sinh qa}\, \psi_{\alpha{\bf p}}(0)-
\frac{\sinh qx}{\sinh qa}\, \psi_{\alpha{\bf p}}(-a).
\label{B-Sol}
\end{eqnarray}
Inserting it into the boundary conditions for the derivatives (\ref{B-derbound}), we have:
\begin{eqnarray}
&&
  \frac{m_B}{m_N}\partial_x\psi_{\alpha{\bf p}}(0)-
  q_0\,\psi_{\alpha{\bf p}}(0)=-
  q_t\,\psi_{\alpha{\bf p}}(-a), 
  \label{boundN}\\
&&
  \frac{m_B}{m_S}\partial_x\psi_{\alpha{\bf p}}(-a)+
  q_0\,\psi_{\alpha{\bf p}}(-a)=
  q_t\,\psi_{\alpha{\bf p}}(0),
  \label{boundS}\\
&&
  q_0=q/\tanh qa,\quad q_t=q/\sinh qa.
  \label{q}  
\end{eqnarray}
These equations serve now as effective 
boundary conditions for the equations in the superconductor and the normal system. For a thick
barrier where $q_t\to 0$, the coupling 
between the "normal" and the "superconducting" operators vanishes, which 
is described by Eqs. (\ref{boundN}) and (\ref{boundS}) 
with zero right-hand sides. 
Using Eqs.~(\ref{boundN}) and (\ref{boundS}), one 
can make sure that the current is continuous at the interface: 

\begin{eqnarray}
 &j_{\alpha {\bf p}}=\frac{i\hbar}{2m_{N}}
  \left[\partial_x\psi^\dag_{\alpha{\bf p}}(0)\psi_{\alpha{\bf p}}(0)-h.c
  \right]=
  \frac{i\hbar}{2m_{S}}
  \left[\partial_x\psi^\dag_{\alpha{\bf p}}(-a)\psi_{\alpha{\bf p}}(-a)-h.c
  \right]=&
\nonumber\\
&=
 \frac{iq_t\hbar}{2m_B}
 \left[\psi^\dag_{\alpha{\bf p}}(0)\psi_{\alpha{\bf p}}(-a)-h.c
 \right].&
\nonumber
\end{eqnarray}

In what follows we assume that the influence of the normal system on the superconductor 
is negligible and the proximity effect in the normal system can be studied without a feedback.
It is normally the case if the superconductor is a bulk metallic system whereas 
the normal system is a degenerate semiconductor where the electron concentration 
is much lower than in the metal. 
In terms of the Fermi momenta this assumption can be expressed as: $p_N\ll p_S$.
Besides we restrict ourselves to the energies $\epsilon$ 
much smaller than the gap energy $\Delta$ in the superconductor: $|\epsilon|\ll \Delta$.
In this case, {\em we will derive closed boundary conditions for the normal electrons which will take into account the proximity-induced electron pairing.}   

It is convenient to include the effective boundary condition at the S side of the barrier (\ref{boundS}) into the equations of motion in the superconductor (\ref{EqS}) by adding appropriate delta-functional terms as follows:

\begin{eqnarray}
&
\left[i\hbar\partial_t-\varepsilon_S({\hat p}_x,{\bf p})-{\hat U}_S(x)+E_S\right]
\psi_{\alpha {\bf p}}(x)
+
\Delta_{\alpha^\prime\alpha}(x)
\psi^\dag_{\alpha^\prime -{\bf p}}(x)=&
\nonumber\\
&=
-\frac{q_t\hbar^2}{2m_B}\delta(x+a)
\psi_{\alpha {\bf p}}(0),& 
\label{EqS1}\\ 
&
\left[-i\hbar\partial_t-\varepsilon_S(-{\hat p}_x,-{\bf p})-{\hat U}_S(x)+E_S\right]
\psi^\dag_{\alpha -{\bf p}}(x)
+
\Delta^*_{\alpha^\prime\alpha}(x)
\psi_{\alpha^\prime {\bf p}}(x)=&
\nonumber\\
&=
-\frac{q_t\hbar^2}{2m_B}\delta(x+a)
\psi^\dag_{\alpha -{\bf p}}(0).&
\nonumber
\end{eqnarray}
where a singular potential
\begin{eqnarray}
{\hat U}_S(x)=\frac{\hbar^2}{2m_B}\delta(x+a)
\left(\partial_x+q_0\right)
\label{US}
\end{eqnarray}
reproduces the boundary condition (\ref{boundS}) with zero right-hand side which 
corresponds to an isolated superconductor. The penetration of Andreev bound 
states in the superconductor at low energies  
is described by a particular solution of Eqs. (\ref{EqS1}) 
generated by the right-hand sides containing the operators
of the N system. It can be expressed in terms of 
the matrix Green function 
of Eqs.~(\ref{EqS1}) whose matrix elements are constructed from 
the quasiparticle
$G^{\alpha\alpha^\prime}_{\bf p}\left(x,t;x^\prime,t^\prime\right)$ and condensate (Gorkov) 
$F^{\alpha\alpha^\prime}_{\bf p}\left(x,t;x^\prime,t^\prime\right)$ Green functions, namely:

\begin{eqnarray}
\left(
\begin{array}{cc}
 \psi_{\alpha {\bf p}}(x,t)\\
 \psi^\dag_{\alpha -{\bf p}}(x,t)
\end{array}
\right)&=&
-\frac{q_t\hbar^2}{2m_B}
\int dt^\prime\times 
\label{solS}\\
&\times & 
\left(
\begin{array}{cc}
G^{\alpha\alpha^\prime}_{\bf p}\left(x,t;-a,t^\prime\right) & 
F^{\alpha\alpha^\prime *}_{-{\bf p}}\left(x,t;-a,t^\prime\right)\\
F^{\alpha\alpha^\prime}_{\bf p}\left(x,t;-a,t^\prime\right) &
G^{\alpha\alpha^\prime *}_{-{\bf p}}\left(x,t;-a,t^\prime\right)
\end{array}
\right)
\left(
\begin{array}{cc}
 \psi_{\alpha^\prime {\bf p}}\left(0,t^\prime\right)\\
 \psi^\dag_{\alpha^\prime -{\bf p}}\left(0,t^\prime\right)
\end{array}
\right).
\nonumber
\end{eqnarray}
These Green functions can be found from Gorkov equations~\cite{Gorkov}
for an isolated superconductor. [The star means complex conjugation].

Inserting solution (\ref{solS}) into the effective 
boundary conditions for $\psi_{\alpha {\bf p}}$  and $\psi^\dag_{\alpha -{\bf p}}$ 
at the N side of the barrier [Eq.~(\ref{boundN})], one finds  
\begin{eqnarray}
&&
\frac{m_B}{m_N}\partial_x\psi_{\alpha{\bf p}}(0,t)-
q_0\,\psi_{\alpha{\bf p}}(0,t)
=
-\frac{q_t^2\hbar^2}{2m_B}\int dt^\prime\times\label{b1}\\
&&
\times\left[ 
G^{\alpha\alpha^\prime}_{\bf p}\left(-a,t;-a,t^\prime\right)
\psi_{\alpha^\prime {\bf p}}\left(0,t^\prime\right)+
F^{\alpha\alpha^\prime *}_{-{\bf p}}\left(-a,t;-a,t^\prime\right)
\psi^\dag_{\alpha^\prime -{\bf p}}\left(0,t^\prime\right)
\right],
\nonumber\\
&&
\frac{m_B}{m_N}\partial_x\psi^\dag_{\alpha -{\bf p}}(0,t)-
q_0\,\psi^\dag_{\alpha -{\bf p}}(0,t)
=
-\frac{q_t^2\hbar^2}{2m_B}\int dt^\prime\times\label{b2}\\
&&
\times\left[ 
G^{\alpha\alpha^\prime *}_{-{\bf p}}\left(-a,t;-a,t^\prime\right)
\psi^\dag_{\alpha^\prime -{\bf p}}\left(0,t^\prime\right)+
F^{\alpha\alpha^\prime}_{\bf p}\left(-a,t;-a,t^\prime\right)
\psi_{\alpha^\prime {\bf p}}\left(0,t^\prime\right)
\right].
\nonumber
\end{eqnarray}
For low-energy states ($|\epsilon|<<\Delta$) these equations 
are reduced to:

\begin{eqnarray}
&&
\frac{m_B}{m_N}\partial_x\psi_{\alpha{\bf p}}(0)-
q_0\,\psi_{\alpha{\bf p}}(0)
=
-\frac{q_t^2\hbar^2}{2m_B}
\left[ 
G\psi_{\alpha {\bf p}}\left(0\right)+
F^*i\sigma_2^{\alpha\alpha^\prime}\psi^\dag_{\alpha^\prime -{\bf p}}\left(0\right)
\right],
\nonumber\\
&&
\label{bound}\\
&&
\frac{m_B}{m_N}\partial_x\psi^\dag_{\alpha -{\bf p}}(0)-
q_0\,\psi^\dag_{\alpha -{\bf p}}(0)
=
-\frac{q_t^2\hbar^2}{2m_B}
\left[ 
G^*\psi^\dag_{\alpha -{\bf p}}\left(0\right)
+Fi\sigma_2^{\alpha\alpha^\prime}\psi_{\alpha^\prime {\bf p}}\left(0\right)
\right],
\nonumber
\end{eqnarray}
where the Green functions are taken at the surface of the superconductor at zero energy
and momentum:
$
G\equiv G(-a;-a)|_{\epsilon ,{\bf p}=0}
$
and 
$
F\equiv F(-a;-a)|_{\epsilon ,{\bf p}=0}.
$
The dependence on the parallel momentum is ignored because for $|{\bf p}|\leq p_N\ll p_S$ 
only electrons moving nearly perpendicular to the interface can
tunnel from the superconductor into the normal system. 

Boundary conditions (\ref{bound}) take into account the conversion of a particle
into a hole (and vice versa) due to Andreev reflection that occurs
simultaneously with normal scattering. 
The advantage of using equations (\ref{bound}) is that the Andreev 
process is described by an explicit coupling between 
the $\psi_{\alpha {\bf p}}(0)$ 
and $\psi^\dag_{-\alpha -{\bf p}}(0)$ operators of the N system with the condensate 
Green function of the superconductor $F$ (and also $G$) reduced to a constant. 
Another advantage is that from the boundary conditions for the operators (\ref{bound})
one can easily derive the boundary conditions for both Green functions and Bogolubov-de Gennes 
functions of the N system.

To study the effect of superconducting proximity on the excitation spectrum of 
a clean N system one needs to take into 
account its finite size determined by the confining potential $V(x)$ in the equations of motion (\ref{EqN}). 
Due to the back wall of the quantum well particles and Andreev reflected holes are 
scattered back to the NS boundary and interfere with those outgoing from the interface,
which generates (in a very subtle way) an energy gap in the quasiparticle spectrum~\cite{Volkov}.  
However in a very narrow well (two-dimensional electron system) 
where the electron states are localized near 
the NS surface the proximity-induced correlations can be treated
in a more robust way. In this case the coupling between 
$\psi_{\alpha {\bf p}}(0)$ and $\psi^\dag_{-\alpha -{\bf p}}(0)$ in the boundary 
conditions (\ref{bound}) represents an effective in-plane pairing for the two-dimensional 
electrons. Indeed, combining equations of motion (\ref{EqN}) and boundary conditions (\ref{bound})
and neglecting all the terms not involving the condensate Green function, one can write: 

\begin{eqnarray}
&
\left[i\hbar\partial_t-\varepsilon_N({\hat p}_x,{\bf p})-V(x)+
E_N\right]\psi_{\alpha {\bf p}}(x,t)=&
\nonumber\\
&
=\delta(x)\left(\frac{q_t\hbar^2}{2m_B}\right)^2
F^{\ast}i\sigma_2^{\alpha\alpha^\prime}
\psi^\dag_{\alpha^\prime -{\bf p}}(x,t),&
\nonumber\\
&
\label{Eq}&\\
&
\left[-i\hbar\partial_t-\varepsilon_N(-{\hat p}_x,-{\bf p})-V(x)+
E_N\right]\psi^\dag_{-\alpha -{\bf p}}(x,t)=&
\nonumber\\
&
=\delta(x)\left(\frac{q_t\hbar^2}{2m_B}\right)^2
Fi\sigma_2^{-\alpha\alpha^\prime}
\psi_{\alpha^\prime {\bf p}}(x,t).&
\nonumber
\end{eqnarray}
Taking into account that for a weakly coupled 2DES 
the transverse wave function $\varphi(x)$ is almost unaffected by the tunneling, 
one can multiply Eqs.~(\ref{Eq}) by $\varphi(x)$ and integrate over $x$ 
(i.e. over the thickness of the 2DES). 
This leads to the equations of motion (\ref{Eq1}) and (\ref{Eq2})
for the operators of the lowest occupied subband with the effective pairing 
energy $E_g$ given by 

\begin{eqnarray}
E_g=\left(\frac{q_t\hbar^2\varphi(0)}{2m_B}\right)^2F.
\label{Eff}
\end{eqnarray}

To estimate $E_g$ one can use the condensate Green function of a superconductor 
with a homogeneous pairing potential 
$\Delta$ at zero energy and parallel momentum: 
$F\approx d^{-1}\sum\nolimits_{p_x}\Delta/(\Delta^2+v_S^2(p_x-p_S)^2)$. 
The integration over the perpendicular momentum $p_x$ gives 
$F\approx 1/\hbar v_S$. 
The boundary value $\varphi(0)$ of the transverse function 
can be estimated using boundary condition 
$\varphi(0)=(m_B/m_Nq)\partial_x\varphi(0)$ [see Eq.~(\ref{boundN})],
where in the right-hand side one can use the "hard wall" wave function
$\varphi(x)=(2/L)^{1/2}\sin\pi x/L$,
which gives $\varphi(0)\approx (m_B/m_Nq)(2/L)^{1/2}(\pi/L)$ 
[$L$ is the thickness of the quantum well].
Thus, the effective pairing energy is  
\begin{eqnarray}
E_g=\frac{\hbar}{m_NLv_S}\,\frac{1}{\sinh^2 qa}\,\frac{\hbar^2\pi^2}{2m_NL^2}\approx 
\frac{v_N}{v_S}\tau E_0,  
\label{Eff_est}
\end{eqnarray}
with $\tau =1/\sinh^2 qa\ll 1$ and the energy of the lowest occupied subband
given by $E_0=\hbar^2\pi^2/2m_NL^2$. Equation (\ref{Eff_est}) is equivalent to 
that obtained in Ref.~\cite{Volkov} for a strong delta-shaped barrier.

\end{document}